\documentclass[twocolumn,showpacs,amsmath,amssymb,superscriptaddress,pre,floatfix]{revtex4}

\usepackage{graphicx}

\begin{document}

\title{Diversity-induced synchronized oscillations in  close-to-threshold excitable elements arranged on regular networks: effects of network  topology}

\author{I. Vragovi\'c}
\email{igor.vragovic@ua.es}
\affiliation{Departamento de F\'isica Aplicada, Instituto Universitario
de  Materiales and Unidad Asociada CSIC-UA, Universidad de Alicante, E-03080 Alicante, Spain}

\author{E. Louis}
\affiliation{Departamento de F\'isica Aplicada, Instituto Universitario
de  Materiales and Unidad Asociada CSIC-UA, Universidad de Alicante, E-03080 Alicante, Spain}

\author{C. Degli Esposti Boschi}
\affiliation{Unit\`a di Ricerca INFM di Bologna,
viale Berti-Pichat 6/2, 40127, Bologna, Italy}

\author{G. Ortega}
\affiliation{Departamento de F\'isica, F.C.E.N. Universidad de Buenos Aires
and CONICET,\\ Pabell\'on I, Ciudad Universitaria, 1428, Capital Federal, Argentine}

\date{\today}

\begin{abstract}
The question of how network topology influences emergent synchronized oscillations in excitable media is addressed. Coupled van der Pol-FitzHugh-Nagumo elements  arranged either on regular rings or on clusters of the square lattice are investigated. Clustered and declustered rings are constructed to have the same number of next-nearest-neighbors (four) and a  number of links twice that of nodes. The systems are chosen to be close-to-threshold, allowing global oscillations to be triggered by a weak diversity among the constituents that, by themselves,
would be non-oscillating. The results clearly illustrate the crucial role played by network topology. In particular we found that network performance (oscillatory behavior and synchronization) is mainly determined by the  network average path length and by the standard deviation of path lengths. The shorter the average path length and the smaller the standard deviation, the better the network performance.  Local properties, as characterized by the clustering coefficient, are less important.
In addition we comment on the mechanisms that sustain synchronized oscillations and on the transient times needed to
reach these stationary regimes.

\vspace{3mm}
\noindent
keywords: excitable media, coupled oscillators, regular networks
\end{abstract}


\maketitle

\section{Introduction}
\label{introduction}

During the last decade, a considerable attention has been drawn to identify 
the mechanisms that may induce global oscillatory behavior in excitable media  
\cite{meron1992,smolen1993,truscott1994,sherman1994,Ca00,hu2000,VS00,VS01,Ha03,AM04,LG04}.
Excitable media are composed of coupled excitable elements characterized by
quiescent states with very large responses (amplitude and relaxation time)
to stimuli larger than some threshold.
This kind of nonlinear behavior has
attracted a great interest in fields such as chemistry \cite{ZZ70,PH97},
and biology \cite{PH97}.
Oscillations in excitable media can be originated by: i) pacemaking cells such as 
in the case of the heart \cite{PH97}, ii) individual elements that oscillate 
when isolated, and, iii) more recently it has been suggested that diversity
amongst the elements and/or internal noise can produce oscillations that are
{\it emergent} in the sense that they show up even when none of
the elements oscillate when isolated.

Motivated by the experimental observation of emergent synchronized oscillations
in the islets of Langerhans \cite{DM68,AS97} (insulin producing structures in
mammalian pancreas) several works have been devoted to offer a sound
explanation for this oscillatory behavior \cite{Ca00,VS00,VS01,EL02}.
Cells in the islets of Langerhans are of the type known as $\beta$ cells and
do not oscillate when isolated \cite{PR91}. As no pacemaking cells are found in
the islets of Langerhans, a different mechanism has to be identified.
In a recent work, Cartwright \cite{Ca00} proposed that emergent synchronized
oscillations may occur in a system of coupled excitable cells or elements in which half of them was 
silent and the other half continuously active, when taken individually.
Each element was described  by means of a van der Pol-Fitzhugh-
Nagumo oscillator \cite{Ca00}. More recently we argued \cite{EL02} that, 
if the system was 
close to threshold, oscillatory behavior could be triggered by just a few diverse
elements. In addition, for large enough coupling 
between the network elements, oscillations may become synchronized.
This proposal is particularly interesting from a biological
point of view as a weak diversity is inherent to biological
systems.

In our previous work we arranged excitable elements  on clusters of the
square lattice. Such periodic two or three dimensional networks have been commonly
used to describe the topology of islets of Langerhans \cite{Ca00,EL02,AF00}
and some types of neurons \cite{PU97}.
The present work is addressed to investigate the role that network topology plays
in the appearance of synchronized oscillations in an excitable medium
described by the model proposed in \cite{Ca00,EL02}.  The effect of topology
on synchronization of oscillators has been intensively investigated in recent
years, particularly in networks having either the small-world or the
scale-free property \cite{Wa99,LH00,BP02,KM02,NM03}. Most studies conclude that the
larger the intrinsic size of the system (quantified by the average network length,
see below) the worse the synchronization. In this work we investigate this issue
on {\it regular} networks with different average path lengths. We choose clusters
of the square lattice with periodic boundary conditions and
rings with additional links such that the degree $k$ (number of
nearest-neighbor connections) equals that of the square lattice clusters ($k$=4).
Our results support the conclusion that the longer the average path length
the worse the synchronization performance. In addition, they indicate
that networks with a small standard deviation of the path length
distribution show better synchronization.
We also discuss the dynamical behavior of the coupled oscillators trying to
identify the mechanisms that trigger and sustain synchronized oscillations.

The paper is organized as follows. In Subsec. \ref{PFHNeq} we discuss the main features of the coupled van der Pol-FitzHugh-Nagumo equations outlining the main reasons why diversity may trigger oscillations. The various measures we use to quantify the degree of oscillatory behavior and synchronization are described in Subsec. \ref{measAS}. However, since we want to deal with (essentially) time-independent indicators, in the Appendix we discuss the stationarity of the present system which is dictated by the relaxation times
to the limit cycles of the oscillators. In order to have a genuine indication of the stationarity, in the appendix we stick
to standard indicators borrowed from circular statistics.  
The topologies of the networks investigated in this work are described in Subsec. \ref{ss_types} as well as the magnitudes used to characterize local and global structure, namely, the average path length and the clustering coefficient. The results are described and discussed in Sec. \ref{subsec_main}. We examine how oscillation and synchronization vary with the coupling between the elements in the network and the effects of the spatial distribution of diverse elements. The differences between networks are highlighted, aiming to identify the characteristics of network topology that determine network performance (as far as synchronization is concerned). We also consider the dependence of global oscillations on the size of the system. Finally, we address the question of how synchronized oscillations are sustained. The achievements of our research are summarized in Sec. \ref{concl}.

\section{Model and Methods}
\subsection{Van der Pol-FitzHugh-Nagumo equations}
\label{PFHNeq}

We describe the excitable medium by means of a system of coupled van der Pol-FitzHugh-Nagumo (PFN) equations \cite{pol1928,hugh1960,high1961,nagumo1962}, arranged on a
given network. The set of equations is written as, \cite{Ca00,EL02,note}:

\begin{subequations}
\label{eq_fhn}
\begin{equation}
\frac{dy_i}{dt} = \gamma \left[ z_i - y_{i}^{3}/3 + y_i + \kappa Q_i \right]
\end{equation}
\begin{equation}
\frac{dz_i}{dt} = - \gamma^{-1} (y_i + \nu_{i} + \beta z_i).
\end{equation}
\end{subequations}
where $i=1,..,N$, $N$ being the total number of elements in the network. Like all relaxation oscillators, PFN oscillator has a variable $y_i$ with a fast release and a variable $z_i$ with a slow accrual phase. The biological
interpretation of these equations is discussed in detail in \cite{Ca00}, here
we briefly comment the main issues. 
Each oscillator is assumed to represent a cell in a living system. 
Variables $y_i$ and $z_i$  correspond to the 
potential across the nonlinear resistance (cell membrane) and the current 
through the supply, respectively. Parameters $\beta$ and $\gamma$ describe the 
(membrane) resistance and the square root of the ratio inductance/capacitance
respectively, and $\nu$
is proportional to the charging potential of the cellular membrane. 

The strength of coupling between nearest-neighbors is denoted by $\kappa$, while the function that describes coupling $Q_i$ is given by,
\begin{equation}
Q_i=\sum_{j=1}^{k_i}(y_j - y_i)=-\sum_{j=1}^{N}L_{ij}y_j \; ,
\label{eq_ct}
\end{equation}
where $k_i$ is the number of nearest-neighbors (or connectivity) of element $i$. 
$L_{ij}$ is usually referred to as the Laplacian matrix, whose structure plays an essential role in determining the behavior of the network, as it contains all the information on network's topology; for instance it has been used to investigate the stability of the synchronized state once established \cite{BP02,NM03}.

In the absence of coupling (isolated element) the parameter $\nu$
determines the behavior of the oscillator. If $\nu$ is outside the oscillating range ($|\nu| > \Theta$, $\Theta$ being the threshold), the element is in stable equilibrium. It is called an excitable element and can be either silent ($\nu < - \Theta$) or continuously active ($\nu > \Theta$). Otherwise, the equilibrium is unstable ($|\nu| < \Theta$) and the element performs oscillations along the limit cycle. The threshold is given by:

\begin{equation}
\Theta = \sqrt{\gamma^2 - \beta} \frac{3 \gamma^2 - 2 \gamma^2 \beta - \beta^2}{3 \gamma^3}
\end{equation}
and for the chosen set of parameters $\gamma=2$ and $\beta=0.5$, kept constant hereafter in this work, it takes the value
$\Theta \cong 0.60412$.

The coupled PFN equations were solved numerically on several regular networks with $N$ nodes, $2N$ links and a connectivity $k_i=4$ for all nodes. All elements are assumed to be active with $\nu_i = \nu$, but a small fraction $x$ of them, called impurities or diverse elements, are taken to be silent with $\nu_i = - \nu$, $i=1,\dots,[xN]$. The numerical results discussed in this work correspond to $\nu=0.61$. Although for this parameter value no element oscillates when isolated, coupling and impurities may trigger oscillations. A simple way to visualize this effect is to write (within a mean field approach) the effective parameter as:

\begin{equation}
\nu^{\rm eff} = \nu (1-2x).
\end{equation}

This clearly shows that the presence of impurities may shift the effective parameter value below the threshold, forcing an originally quiescent element to oscillate. Although this expression can be used to estimate the fraction of impurities above which the oscillatory behavior emerges, it cannot account for the important differences in the performance of ordinary and impurity elements that can actually show up (see below). Fortunately, an effective $\nu$ can be estimated separately for each element \cite{Ca00}:
\begin{equation}
\nu_{i}^{\rm eff} = \nu_i - \beta \kappa \sum_{j=1}^{k_i}(y_j - y_i) .
\label{eq_nueff1}
\end{equation}
At quiescence there is no coupling between elements having the same parameter value. However, in the case of one silent ($\nu_j = - \nu$) and one continuously active cell ($\nu_i = + \nu$) we can write $(y_j - y_i) \approx - (\nu_j - \nu_i) = 2 \nu$. The coupling between different cells is therefore of crucial importance for the effective change of cells parameters. Let us call $m_{\pm}(x)$ the average number of the nearest-neighbors of opposite nature for ordinary ($+$) and impurity ($-$) cells. The effective parameter can then be written as:

\begin{equation}
\nu_i^{\rm eff} = \pm \nu \left[ 1 - 2 m_{\pm}(x) \beta \kappa  \right] .
\label{eq_nueff2}
\end{equation}

\noindent For large enough $N$, the average number of nearest neighbors $m_{\pm}(x)$ is given by,

\begin{equation}
m_{+}(x) = \sum_{n=1}^{4} n \binom{4}{n} x^n (1-x)^{4-n}=4x \; ,
\label{eq_nueff3}
\end{equation}

\begin{equation}
m_{-}(x) = \sum_{n=1}^{4} n \binom{4}{n} x^{4-n} (1-x)^{n}=4(1-x) \; .
\label{eq_nueff4} 
\end{equation}

This indicates that, on average, impurities can be much more affected by coupling than ordinary cells. To get a grasp of the implications of this result, let us give some numbers. The critical fraction of impurities $x_{\rm c}$ required to force ordinary cells to oscillate can be derived by setting $\nu_{+}^{\rm eff} = \Theta$:
\begin{equation}
x_{\rm c} = \frac{1}{8 \beta \kappa} \left(1 - \frac{\Theta}{\nu}\right).
\end{equation}
Taking a typical value for the coupling strength $\kappa=1$ and the values of the other two parameters given above, we obtain 
$x_{\rm c} \cong 0.00241$. The effective parameter of the impurity cell for this value of $x$ is $\nu_{-}^{\rm eff} \approx + 1.8$. The effect is so strong that the parameter has not only changed its sign, but it is also well outside the oscillatory regime. This result underlies the differences in the behavior of ordinary and impurity cells discussed below.

\subsection{Measures of oscillation and synchronization}
\label{measAS}

We quantify the emergence of oscillatory behavior by calculating various averages
and standard deviations, over a sufficiently long time interval (see below), of either the fast variable $y_i$ or the phase,
both for individual elements and for the network as a whole.
In the literature one can find many definitions of synchronization, mainly because of the different physical meanings
(full-, phase-, delayed-synchronization, etc.)
Measures similar to the ones adopted here are commonly used to 
characterize oscillatory behavior in excitable media 
\cite{Ca00,Ha03,EL02,LH00,GH01,HC02,KH01,MH03}.
In order to make the bridge with well-established circular statistics \cite{mardia},
in the Appendix we evaluate the circular variance for some selected cases, drawing the same conclusions (at least
qualitatively) of this section as far as the level of synchronization is concerned. 
As a byproduct we also obtain information on the stationarity of our model.

As discussed elsewhere \cite{EL02} transients can be quite long for weakly cells. Instead, in the case of strongly coupled cells the coherent
oscillatory behavior is almost instantaneously attained (cp. Fig. 1 in Ref. \cite{EL02}, see also Appendix). In order to minimize the contribution of transients \cite{EL02}, the averages are computed starting from a large initial time $t_{\rm i}$ after which the stationary state is reached (for moderate and strong coupling regimes), and over a time interval $\Delta t = t_{\rm f}-t_{\rm i}$ long enough to cover a sufficient number of periods. In a typical case, the coupled PFN
equations were solved with an integration step of $0.05$, taking averages from $t_{\rm i}=200$ up to $t_{\rm f}=400$. The initial time $t_{\rm i}=200$ is large enough to surpass the transient period of moderately coupled networks with low synchronization (see Appendix). For the values of cell parameters given above, the intrinsic period is around $T=10$, so that oscillations were averaged over approximately 20 periods.

We use the following standard deviation as a measure of intensity of cell oscillations (CO) at node $j$:

\begin{equation}
\sigma^{\rm CO}_{j}=\sqrt{\frac{1}{\Delta t} \sum_{t=t_{\rm i}}^{t_{\rm f}} \left[ y_j^2(t)-\langle y_j \rangle^2 \right]},
\end{equation}
where $\langle y_j \rangle = 1/\Delta t \sum_{t_{\rm i}}^{t_{\rm f}} y_j(t)$ is the time average over interval $\Delta t$. A high value of $\sigma^{\rm CO}_{j}$ implies a large amplitude of cell oscillations around its time average, while a low value reveals an almost non-oscillatory behavior.

We also define the average of cell oscillations over the whole network as:

\begin{equation}
\sigma^{\rm CO}=\frac{1}{N} \sum_{j=1}^{N}  \sigma^{\rm CO}_{j} .
\label{eq_sCA2}
\end{equation}

Depending on the neighborhood, the effective parameters $\nu_{j}^{\rm eff}$ can largely differ from cell to cell (see preceding subsec.), inducing different oscillation intensities. The dispersion of cells oscillation intensity with respect to its network average can be quantified by means of the standard deviation:

\begin{equation}
\sigma^{\rm DO}=\sqrt{\frac{1}{N-1} \sum_{j=1}^{N} \left[ (\sigma^{\rm CO}_{j})^2-(\sigma^{\rm CO})^2 \right] }.
\label{eq_sDA}
\end{equation}

Synchronization is evaluated by comparing the current phases of each oscillator with the current phase of a randomly chosen reference cell (ordinary or impurity), and calculating the time average over the interval $\Delta t$ \cite{EL02}:

\begin{equation}
\sigma^{\rm S}=\sqrt{\frac{1}{(N-1) \Delta t} \sum_{j \neq r}^{N} \sum_{t=t_{\rm i}}^{t_{\rm f}} \left[ \cos \phi_i(t)- \cos  \phi_{r}(t) \right]^2 }.
\label{eq_sS}
\end{equation}
Shifting all the limit cycles to have a common center we have defined
\begin{equation}
\cos \phi_i(t)=\tilde{y}_j(t)/r_j(t)= \tilde{y}_j(t)/\sqrt{\tilde{y}_j(t)^2+\tilde{z}_j(t)^2}
\label{phi}
\end{equation}
where $\tilde{y}_j(t)=y_j(t)-\langle y_j(t) \rangle$ and
$\tilde{z}_j(t)=z_j(t)-\langle z_j(t) \rangle$. The sum in $\sigma^{\rm S}$ is not extended to all possible reference sites in order to limit the computational time. However, singling out a reference site allows us to track the difference in the behavior
of ordinary and diverse elements while, for instance, this information is hidden in a measure like the circular
variance that makes reference to an average phase (see Appendix).
The smaller $\sigma^{\rm S}$ the better the synchronization. The measure chosen to test synchronization in this work is more demanding than those used in previous works, in the sense that now we evaluate the differences in both dynamical variables \cite{Ca00,EL02}. Note that the  average values are subtracted from each dynamical variable in  in Eq. (\ref{eq_sS}). This means that this measure does not account for possible differences in these averages, that can arise from different effective parameters $\nu^{\rm eff}$.

In order to improve synchronization, first the majority of cells should be oscillating with the same frequency. Once that is achieved, dephasing between cells should vanish. Thus, an essential parameter for characterizing the emergence of synchronization is the dispersion of oscillating periods. We quantify this dispersion by means of the standard deviation divided by the average period:

\begin{equation}
\sigma^{\rm DP}=\frac{1}{\bar T}\sqrt{\frac{1}{N-1} \sum_{j=1}^{N} \left[ T(j)^2 - \bar{T}^2 \right] },
\label{eq_sDP}
\end{equation}
where $\bar{T}=1/N \sum_{j=1}^{N} T(j)$ is averaged over the whole network.

Finally, we can measure the intensity of oscillations the network as a whole comparing the time dependent network oscillations with the temporal average over the interval $\Delta t$ \cite{LH00}.

\begin{equation}
\sigma^{\rm NO}=\sqrt{\frac{1}{\Delta t} \sum_{t_{\rm i}}^{t_{\rm f}} \left[ \bar{y}(t)^2-\langle \bar{y} \rangle^2 \right]},
\label{eq_sNA}
\end{equation}
with the network activity $\bar{y}(t)=1/N \sum_{j=1}^{N} y_j(t)$ and its temporal average $\langle \bar{y} \rangle = 1/\Delta t \sum_{t_{\rm i}}^{t_{\rm f}} \bar{y}(t)$.
In order to have large values of $\sigma^{\rm NO}$, the majority of cells must both oscillate and be synchronized. Thus, the intensity of the network oscillations $\sigma^{\rm NO}$ is an appropriate quantity for the simultaneous measure of both activation and synchronization in excitable media \cite{LH00}.

\subsection{Networks}
\label{ss_types}

\begin{figure}[h]
\includegraphics[angle=0,width=9cm]{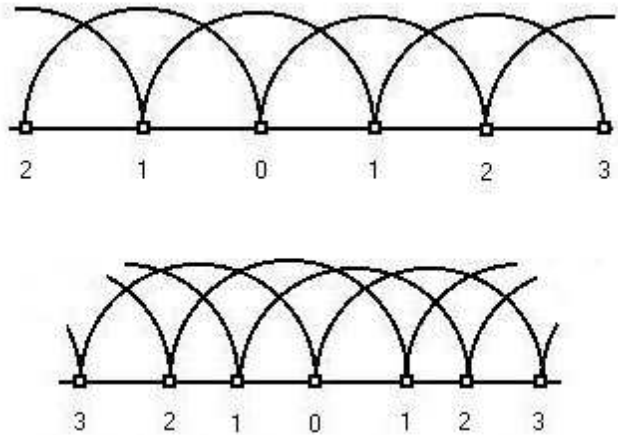}
\caption{Illustrates the link structure in clustered rings (upper) with
connections to the first and
second nearest-neighbors ($K=2$), and declustered rings (lower) with
connections to the first and third nearest-neighbors ($K=\bar{3}$). Note that while the
clustered ring has triangles of elements the declustered ring has quadrilaterals.}
\label{rings}
\end{figure}

As the main aim of this work is to investigate the role of topology, we shall compare the results obtained on various regular networks having a number of links twice that of nodes and the same degree (number of nearest neighbors). We choose the square lattice with periodic boundary conditions (this guarantees that all nodes $i$ have degree $k_i=4$) and two types of regular rings (see Fig. \ref{rings}). The first one is the simplest clustered regular lattice, which consists of a ring with additional connections to the two next-nearest neighbors (this is usually referred as $K=2$). The second type of a ring has zero clustering (see below) and is constructed as follows. Standard (clustered) rings with $K=n$ are built by linking each site to all neighbors from the first up to the $n$-th. Declustered rings (see \cite{VL05}), denoted as $K={\bar n}$, have additional links only to the $n$-th neighbors. In this work we investigate rings with $K=2$ and $K={\bar 3}$, both having degree $k_i$=4 for all nodes.

The topology of connections between sites can strongly influence the cooperation and the exchange of information between them. The structural properties of a graph are usually quantified by the average path length $L$ and the clustering coefficient $C$ \cite{wattsstrogatz1998,albertbarabasi2001,VL05}. The average path length is calculated as the network average of the shortest graph distances between two nodes for all possible pairs,
\begin{equation}
L = \frac{1}{N(N-1)} \sum_{i \neq j} d_{ij}.
\end{equation}

The clustering coefficient $C$ measures to which extent are the neighbors of each site connected to each other. It is defined as follows,
\begin{equation}
C = \frac{1}{N} \sum_{i} \sum_{j \neq k}\frac{n_{jk}}{k_i(k_i-1)/2},
\end{equation}
\noindent where $j$ and $k$ are nearest-neighbors of $i$ and $n_{jk}=1$ only if there is a link betwee $j$ and $k$. This definition does only give $C\neq 0$ if there are triangles in the network. This is the case of clustered rings such as that chosen here ($K=2$) but it is not that of the square lattice nor  that of the declustered ring, both having $C=0$.

The average length for the three networks investigated here can be calculated analytically. The result for the clustered ring is,
\begin{equation}
L(K=2) = \frac{N}{4}\frac{N/2+1}{N-1} \; ,
\end{equation}
\noindent valid only for $N=4n$, where $n$ is an integer. For very large systems this behaves as $L \sim N/8$. In the case of the declustered ring we obtain,
\begin{equation}
L(K={\bar 3}) = \frac{N}{6}\frac{N/2+4}{N-1} \; ,
\label{apl_b3}
\end{equation}
\noindent valid only for networks with a number of nodes $N=6n$, $n$ being an integer. In this case the limit for $N \rightarrow \infty$ is
$L \sim N/12$.

It is interesting to write a general formula for an arbitrary $K$ in the case of a declustered ring,
\begin{equation}
L(K) = \frac{N}{2K}\frac{N/2+(K^2-i)/2}{N-1} \; ,
\end{equation}
\noindent where $i$=1,2 for $K$ odd or even, respectively. This formula is valid for $N=2Kn$, where $n$ is an integer.

Finally, the result for clusters of the  square lattice with periodic boundary conditions is,
\begin{equation}
L({\rm square}) = \frac{N^{3/2}}{2(N-1)} \; ,
\label{apl_sq}
\end{equation}
\noindent its asymptotic behavior being $L \sim 0.5\sqrt{N}$.

The main difference between rings (clustered or declustered) and the square lattice is that the latter is far smaller than the former as its average length increases sublinearly with the network size. It is worth trying to understand why the square lattice behaves in this way in terms of the structure of the Laplacian matrix.
By appropriately renumbering sites, any cluster of the square lattice with periodic boundary conditions can be mapped onto a ring in which additional links are introduced. These links are: links between
$\sqrt{N}$ nodes and their $(\sqrt{N}-1)$-th and $(\sqrt{N}+1)$-th neighbors, and
links between the remaining $N-\sqrt{N}$ and their $(\sqrt{N}+1)$-th neighbors.
For large $N$ this is equivalent to a declustered ring with $K=\bar{\sqrt{N}}$. In fact Eq. (\ref{apl_b3}) for declustered rings gives, for $K=\bar{\sqrt{N}}$, the same assymptotic behavior than that obtained for  clusters of the square lattice (see Eq. (\ref{apl_sq})).  Contrary to regular rings with $K=2$ or $K=\bar{3}$ where additional links introduce non-zero elements in the second and third subdiagonals of the Laplacian matrix, respectively, links in the square lattice are long-ranged and introduce non-zero elements in the $(\sqrt{N}-1)$-th and $(\sqrt{N}+1)$-th subdiagonals.

\section{Results}
\label{subsec_main}
Results discussed hereafter were obtained for the following values of the model
parameters: $\gamma=2$ and $\beta=0.5$ were kept constant, while we took $\nu=0.61$
for normal elements and $\nu=-0.61$ for diverse elements. The concentration of impurities is 2\%. The number of elements was mainly fixed at $N=100$ although we also consider
clusters of the square lattice of size $18\times18$. In all cases
initial conditions were $y_i(0)=1$ and $z_i(0)=1$ for all network nodes $i$.
We have previously checked that varying the initial data within
the limit cycle has a very small effect on the indicators of Subsec. \ref{measAS}
evaluated after the transients.

\begin{figure}[h]
\includegraphics[angle=270,width=8cm]{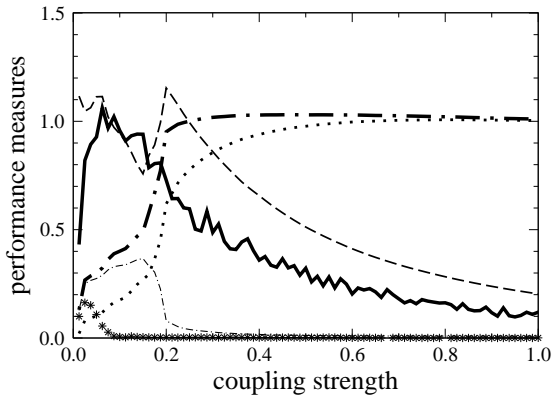}
\caption{Measures used to characterize the emergence of oscillations and synchronization versus strength of coupling between excitable elements in the PFN system described by Eqs. (\ref{eq_fhn}). Excitable elements are arranged on a square network of linear size $l=10$, with 98 continuously active sites ($\nu=0.61$) and 2 silent impurities ($\nu= - 0.61$). The results are averaged over 10 different realizations of impurity distributions. Cells oscillation ($\sigma^{\rm CO}$, see
Eq. (\ref{eq_sCA2})) - thick dot-dashed line, network oscillations  ($\sigma^{\rm NO}$ of
Eq. (\ref{eq_sNA})) - dotted, and synchronization ($\sigma^{\rm S}$, see Eq. (\ref{eq_sS})) with respect
to an ordinary (full), or an impurity element (dashed). Dispersion of element
oscillation ($\sigma^{\rm DO}$, of Eq. (\ref{eq_sDA})) - thin dot-dashed, and  of the
oscillation periods ($\sigma^{\rm DP}$, see Eq. (\ref{eq_sDP})) - stars (see text)
are also shown.}
\label{squareK}
\end{figure}

\subsection{Coupling Strength}
\label{ss_cs}

The analysis of the influence of coupling between cells on intensity of oscillations and synchronization reveals three different regimes (see Fig. \ref{squareK}). The results shown in the Figure correspond to a cluster of the square lattice having $N=100$ nodes with a fraction of impurities $x=0.02$.

In the case of weak coupling ($\kappa < 0.1$), the chosen transient $t_{\rm i}=200$
is not long enough to allow the system to reach the stationary state.
The behavior of the system is strongly influenced by the initial conditions.
The improvement of synchronization as coupling strength decreases to zero is a
consequence of the initial conditions (common to all elements). Increasing $t_{\rm i}$
would allow the weakly coupled oscillators to run away from each other, see Appendix, deteriorating synchronization. Only after rather long $t_{\rm i}$ the network
will reach the stationary oscillating state. As $\kappa$ goes to zero the
transient period will become infinite. Results shown in Fig. \ref{squareK}
for such low values of coupling cannot be considered too reliable.

\begin{figure}[h]
\includegraphics[angle=270,width=8cm]{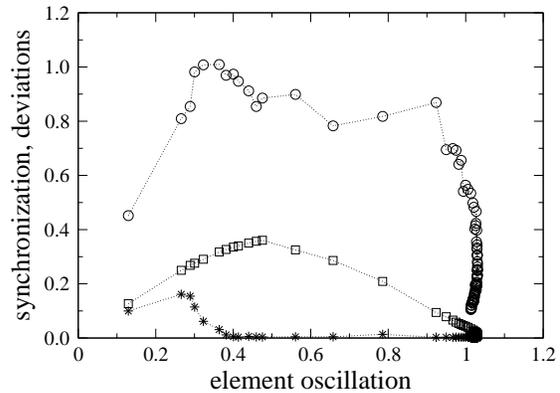}
\caption{Synchronization (circles) $\sigma^{\rm S}$ of Eq. (\ref{eq_sS}), dispersion of oscillation (squares) $\sigma^{\rm DO}$ of Eq.
(\ref{eq_sDA}), and dispersion of periods (stars) $\sigma^{\rm DP}$ of Eq. (\ref{eq_sDP}), versus cell oscillation $\sigma^{\rm CO}$
of Eq. (\ref{eq_sCA2}). Elements were arranged on clusters of the square lattice with $N=100$ with 98 continuously active sites ($\nu=0.61$) and 2 silent impurities ($\nu= - 0.61$). The results are averaged over 10 different realizations of impurity distributions. Coupling strength is varied over the range $\kappa \in (0,1)$.}
\label{squareactsyndevK}
\end{figure}

At moderate coupling ($0.1 < \kappa < 0.3$), the perturbation originating from impurity cells is still not spread much over the system. Ordinary cells are only weakly impelled to oscillate, but with different amplitudes. The dispersion of
oscillations is large, reaching its maximum for
$\kappa \approx 0.15$, cp. Fig. \ref{squareK}. In addition, a moderate coupling is not strong enough to eliminate large dephasing between elements oscillating with the same frequency. Consequently, the oscillation of the network as a whole is much weaker than that of individual elements, as synchronization is still not attained.

In the strong coupling regime ($\kappa > 0.3$), oscillations of elements become more and more intensive and synchronization is gradually improved, leading to a strong oscillation of the network as a whole. Elements intensively oscillate and there are no significant differences in frequencies (the dispersion of periods is practically zero). Increasing the coupling strength further, the phase differences are eliminated and synchronization is improved. Finally, ordinary elements are much better synchronized with each other, while the impurity elements are forced to oscillate in the same way as the rest of the network only in the case of very strong coupling, cp. Fig. \ref{squareK}. Reasons for the different behavior of ordinary and impurity elements were given in Subsec. II.A.

\begin{figure}[h]
\includegraphics[angle=270,width=8cm]{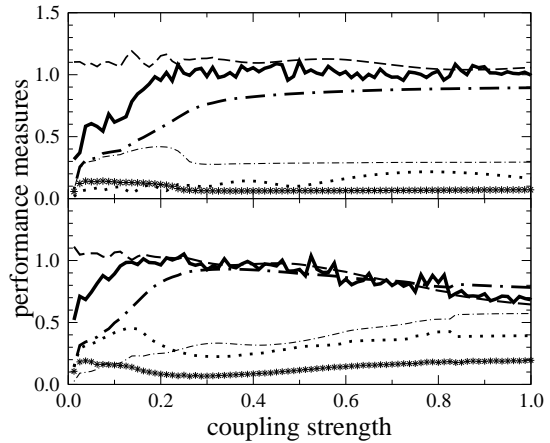}
\caption{Same as in Fig. \ref{squareK} but with different network topologies: Excitable elements are arranged
either on clustered (upper) or declustered (lower) rings both with $N$=100 nodes,
with 98 continuously active sites ($\nu=0.61$) and 2 silent impurities ($\nu= - 0.61$).}
\label{regKc}
\end{figure}

The aforementioned mechanism of improving synchronization is even more evident if we plot intensity of oscillations, network synchronization, and dispersion of intensities and period of oscillations in a single graph (Fig. \ref{squareactsyndevK}). Synchronization is the worst when the deviation of periods reaches its maximum for $\sigma^{\rm CO}$ around $0.3$. After that, it is practically unchanged until all elements oscillate with the largest amplitude. Approaching the maximal activation, the dispersion of oscillations decreases, and synchronization is rapidly improved by eliminating differences in phases.

In the case of rings, the values of coupling strengths needed both to shorten the transient period below a given $t_{\rm i}$ and to synchronize the elements are larger, cp. Fig. \ref {regKc}. The long average path length of the clustered ring ($L=12.88$ to be compared with $L=5.05$ for square lattice of the same size), hinders the effective spread of perturbations coming from impurity elements.  Both the dispersion of oscillations and the dispersion of periods remain large. The network reaches
the steady regime for $\kappa > 0.5$. Similar results are obtained for declustered rings, with a modest improvement of $\sigma^{\rm S}$. Oscillation and synchronization are again quite low, due to the large average path length ($L=9.09$), cp. Fig. \ref {regKc}. This  indicates that  collective oscillatory behavior is determined mainly by the average length of the network, and not by its local properties quantified by the clustering coefficient.
Note that the square network and the declustered ring, both having zero clustering and appeciably different average path lengths, have very different performances. The shorter the length the better the performance.

\begin{figure}[h]
\includegraphics[angle=270,width=8cm]{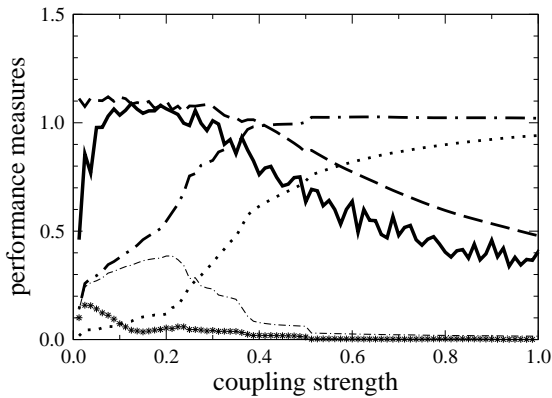}
\caption{Same as in Fig. \ref{squareK} but with a different amount of diversity: Excitable elements are arranged on a square network of linear size $l=18$, with 318 continuously active sites ($\nu=0.61$) and 6 silent impurities ($\nu= - 0.61$).}
\label{fig18sq}
\end{figure}

\begin{figure}[h]
\includegraphics[angle=270,width=8cm]{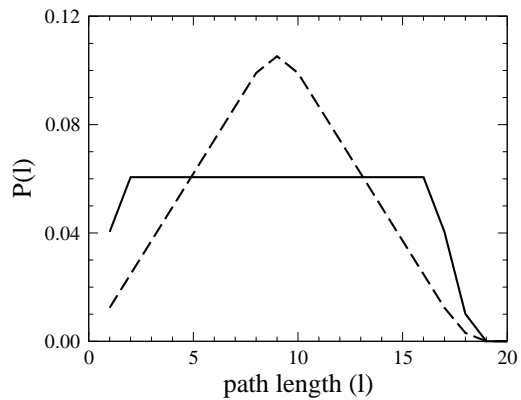}
\caption{Probability density function of path lengths in a declustered ring with $N$=100 nodes and $K={\bar 3}$ (continuous line) and in a cluster of the square lattice with $N=18^2$ nodes (broken line).}
\label{fig:prob}
\end{figure}

Now, the question is whether the average path length is or not the only topological feature that matters. We address this question analyzing a cluster of the square lattice of a bigger size (18$\times$18, with $N=324$), having almost the same characteristic length ($L=9.03$) as a declustered ring with $K={\bar 3}$ and 100 sites (see above). The coupling strengths are varied over the same range ($\kappa \in (0,1)$), while the fraction of impurities is not changed ($x$=0.02 as above). As shown in Fig. \ref{fig18sq} performance is now much better than that found for the declustered ring (see Fig. \ref{regKc}) indicating that the average length is not enough to characterize network behavior. Fig. \ref{fig:prob} shows the probability density function of path lengths in these networks. The density shows a well defined peak at the average path length in the case of the square lattice cluster, while it is almost constant in the declustered ring. As a consequence the standard deviation of path lengths is appreciably smaller in the former (3.7 to be compared with 4.9 for the declustered ring). We believe that the different standard deviations of path lengths is the origin of the noticeably different performance of the two networks. This result suggest that more homogeneous systems, as far as  path length is concerned, are more efficient in triggering collective oscillations and synchronization.

\subsection{Distance between diverse elements}
\label{ss_di}

\begin{figure}[h]
\includegraphics[angle=270,width=8cm]{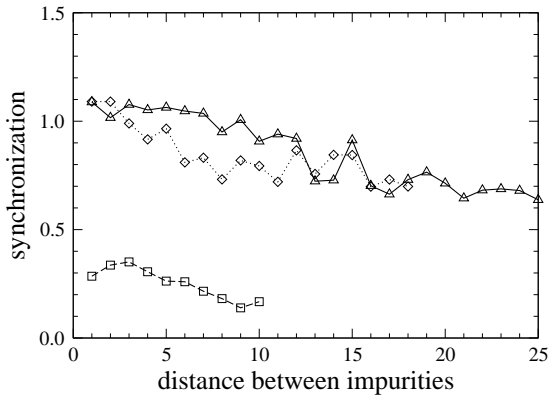}
\caption{Synchronization (ordinary reference site) versus distance between the two impurity elements in several networks of size $N=100$: square network (square), clustered (triangle) and declustered (diamond) rings. Coupling strength and fraction of impurities are $\kappa=0.5$ and $x=0.02$, respectively.}
\label{compS}
\end{figure}

All the results presented up to now are averaged over several realizations of impurity distribution. It turns out that the oscillating behavior strongly depends on the distance between impurity elements. It is likely that larger distance between them would improve the performance, as the overlapping of their respective influence domains is smaller. In other words, the average path length from ordinary sites to the closest impurities becomes shorter. This is illustrated in Fig. \ref{compS} where synchronization is plotted as a function of the distance between impurities.  Rings and square clusters behave in a similar way (the pace at which synchronization improves with distance is  similar in all networks) although synchronization is much poorer in the former.

\subsection{The Size of the Network}
\label{ss_sn}

We have investigated how synchronization varies with the network size in the case of the square lattice. Calculations were carried out keeping the fraction of impurities constant $x=0.02$. As argued in \cite{VS01} the effective coupling constant varies with the average network length as $1/L^2$. This can be easily checked by noting that in the continuum limit the coupling term of Eq. (\ref{eq_ct}) behaves as,
\begin{equation}
Q \propto \frac{\partial^2 y}{\partial \eta^2},\;\;\;\;\;\;\;\;\;  0 < \eta < L.
\end{equation}
Then, rescaling the spatial variable $\eta$ to bring to evidence size effects we find,
\begin{equation}
Q \propto \frac{1}{L^2}\frac{\partial^2 y}{\partial \eta^2},\;\;\;\;\;\;\;\;  0 < \eta < 1,
\end{equation}
\noindent where $L$ should be interpreted as the average linear size of the system.

\begin{figure}[h]
\includegraphics[angle=270,width=8cm]{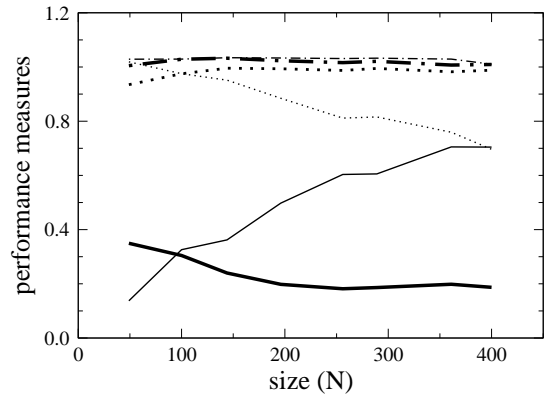}
\caption{Element oscillation (dot-dashed line), network oscillation (dotted line) and synchronization with respect of an ordinary reference site (full line) versus size of clusters of the square lattice. The coupling strength was taken either constant ($\kappa=0.5$, thin lines) or proportional to the square of the average path length ($\kappa=0.5 N/100$, thick lines).}
\label{size2D}
\end{figure}

In order to check this behavior we have carried out calculations with a coupling strength  either constant ($\kappa=0.5$)  or proportional to $L^2$ ($\kappa=0.5N/100$). The numerical results are depicted in Fig. \ref{size2D}. We first note that, for $\kappa=0.5$, synchronization is gradually deteriorated as the size of the system increases. We have checked that the chosen transient $t_{\rm i}=200$ is long enough for the largest network shown in the Figure $N=400$. The pace at which it worsens is rather slow due to the sublinear behavior of the average path length (see above). Rescaling the coupling strength as just mentioned, gives a synchronization that remains almost unchanged as the size increases, proving the validity of the scaling argument. A similar analysis for rings shows a much faster worsening of the system performance when coupling is kept constant, again due to the fast (linear) increase of the average length with the network size.

\subsection{Sustaining synchronized oscillations}

\begin{figure}[h]
 \includegraphics[angle=270,width=8cm]{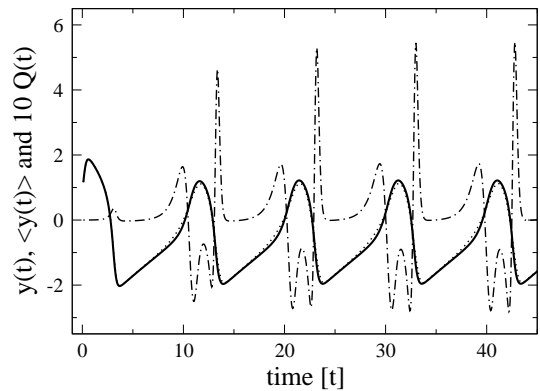}
\caption{Time dependence of  oscillation of a randomly chosen ordinary element ($y_i(t)$ - full line), its coupling function ($10 \times Q_i(t)$ - dot-dashed) and oscillation averaged over the whole network ($\langle y(t) \rangle$ - dots almost superimposed to $y_i(t)$).
Elements were arranged on a square cluster of linear size $l=10$. The results correspond to a single realization of impurity distribution. The chosen coupling strength ($\kappa=0.5$) is large enough to trigger synchronized oscillations.}
\label{syndif}
\end{figure}

The results and discussion of the preceding subsections clearly show that coupling between elements and diversity are the main cause of synchronized oscillations in the systems under study. Coupling is finite only if $y_i$ are not all identical. Then, how is synchronization sustained?  Fig. \ref{syndif} shows  $y_i(t)$ and the coupling function $Q_i(t)$ for a single element $i$ as a function of time. The average over the whole network $\langle y(t) \rangle$ is also shown. The results correspond to the strong coupling regime ($\kappa=0.5$). First we note that a characteristic oscillating pattern and coherent behavior is achieved almost instantaneously. We have checked that
this is the case even if the initial conditions are varied. On the other hand no significant differences between oscillation of a particular element and of the network as a whole are noted. The coupling function $Q_i(t)$ shows an oscillatory pattern similar to that of  $y_i$, although it is an order of magnitude smaller. This shows that a small difference between variables in each element is enough to sustain synchronized oscillations. In addition the oscillatory pattern is stable: any
perturbation of phase coordinates (within the range of limit circle) at time $t_0$, could be viewed as setting new initial conditions starting at that time. The effect of the perturbation would then be eliminated after a few periods, restoring the characteristic pattern.

\section{Conclusions}
\label{concl}

The conclusions that emerge from our study of the induction of global synchronized oscillations by heterogeneity in excitable regular systems are:

1. Elements and network activity and synchronization, as collective phenomena, depend mainly on global properties of the network. Shortening the average path length by merely changing the topology, while keeping constant the number of sites and links, improves synchronization. In addition we found that the standard deviation of  path lengths is also a very crucial variable: the smaller it is the better the network performance becomes. This indicates that homogeneity, as related to path length, is a benefitial feature of networks.

2. Local properties quantified by the clustering coefficient do play a role much less significant than the average path length and its standard deviation.

3. Synchronization in clusters of the square lattice is improved by increasing the strength of coupling between the excitable elements. In the strong coupling regime most of the elements are highly activated, oscillating with the same frequencies and negligible dephasing.

4. Circular regular networks with short range links are hardly synchronized even when elements are strongly coupled, due to their long average path length and large standard deviation.

5. Oscillation and  synchronization depend on the spatial distribution of impurity elements, improving as the distance between those elements increases.

6. Finally, we have revisited the behavior of van der Pol-FitzHugh-Nagumo excitable media when their size is varied. We found that the dynamical behavior of the network can be kept unchanged when the system size increases, provided that the coupling strength is increased proportionally to the square of the
average path length. Regarding synchronization, the optimal topologies are the ones for which the average path length increases more slowly with the size of the system.

\begin{acknowledgments}
Financial support by Fet Open Project COSIN IST-2001-33555 and the Universities
of Barcelona and Alicante is gratefully acknowledged.
\end{acknowledgments}

\appendix
\section{Circular variance and stationarity}
\label{app_cv}

The degree of synchronization of the coupled PFN oscillators as well as its dependence on time can be analyzed using
well-established measures of circular statistics \cite{mardia}.
Given the set of phases $\phi_i(t)$ of Eq. (\ref{phi}) as a function of time and
the corresponding
unit vectors $\widehat{OP_i}$, the average phase $\Phi(t)$ is defined to be the
direction of the resultant $\widehat{OP}$. The cartesian coordinates of $P$ are:

\begin{equation}
\bar{C}(t) = \frac{1}{N} \sum_{i=1}^{N} \cos \phi_i(t) , \hspace{2mm}
\bar{S}(t) = \frac{1}{N} \sum_{i=1}^{N} \sin \phi_i(t),
\end{equation}
leading to the length of the resultant:
\begin{equation}
R(t) = N \bar{R}(t) = N \sqrt{\bar{C}(t)^2 + \bar{S}(t)^2}.
\end{equation}

The average phase $\Phi(t)$, i.e. the mean direction, is obtained from the
following equations:

\begin{equation}
\bar{C}(t) = \bar{R}(t) \cos \Phi(t) , \hspace{2mm}
\bar{S}(t) = \bar{R}(t) \sin \Phi(t).
\end{equation}

Recalculating the current phases relative to the mean
($\xi(t) = \phi(t) - \Phi(t)$), one can see that the sum of deviations
about the mean equals zero:

\begin{equation}
\sum_{i=1}^{N} \sin \xi_i(t) = \sum_{i=1}^{N} \sin (\phi_i(t) - \Phi(t)) = 0.
\end{equation}

As a possible measure of the synchronization one could use the {\it circular variance} \cite{mardia},
given by:
\begin{equation}
D(t) = \frac{1}{N} \sum_{i=1}^{N} [1 - \cos (\phi_i(t) - \Phi(t))].
\end{equation}
It can be shown that the variance $D$ is minimized, if the phases are
recalculated relative to the mean \cite{mardia}.

\begin{figure}[h]
\includegraphics[angle=270,width=8cm]{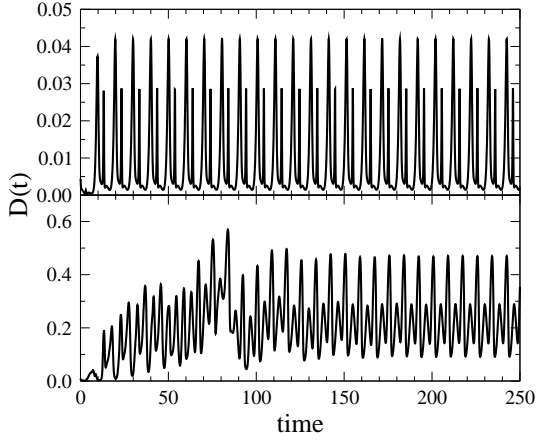}
\caption{Time dependent circular variance $D(t)$ for $N=100$ oscillators with two impurities
coupled into a square lattice with strong and moderate interaction. Upper: $\kappa = 0.5$; Lower: $\kappa = 0.125$.}
\label{a12}
\end{figure}

Small values of  $D$ indicate that the phases of coupled oscillators
are grouped around the mean, hence oscillators being well synchronized. On the
other hand, large values of $D$ correspond to rather uniform distribution of
phases and low synchronization.

\begin{figure}[h]
\includegraphics[angle=270,width=9cm]{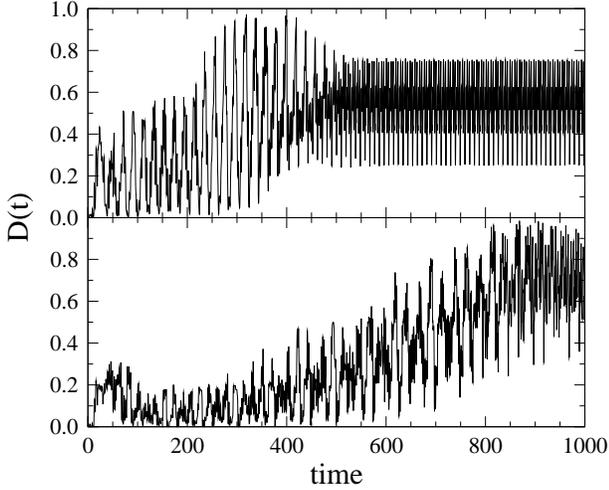}
\caption{Time dependent circular variance $D(t)$ for $N=100$ oscillators with two impurities
coupled into a square lattice with small interactions. Upper: $\kappa = 0.025$;
Lower: $\kappa = 0.0125$.}
\label{a34}
\end{figure}

Let us analyze first the square lattice. In the case of strong coupling ($\kappa=0.5$) oscillators are well synchronized, with $D(t)$ having very small values, cp. Fig. \ref{a12} (upper). Better synchronization is achieved in the slower part of the limit cycle, while oscillators tend to run away from each other in the faster part of the path, increasing $D(t)$ towards 0.045. The coupling is strong enough to lead to stationary behavior already after 2 or 3 periods.

\begin{figure}[h]
\includegraphics[angle=270,width=9cm]{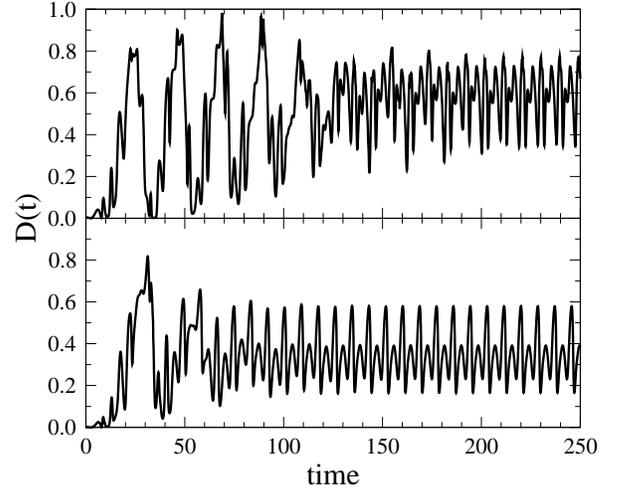}
\caption{Time dependent circular variance $D(t)$ for $N=100$ oscillators coupled into a regular ring with two impurities and $\kappa = 0.5$. Upper: clustered ring, $K=2$; Lower: declustered ring, $K=\bar{3}$.}
\label{a56}
\end{figure}

Decreasing the coupling strength, deteriorates synchronization. In the case of $\kappa=0.125$, the variance $D(t)$ takes much larger values (up to 0.5 in stationary regime), while the transient extends to $t=125$.
Decreasing further  the coupling strength ($\kappa = 0.025$) worsens the synchronization and increases the transient period ($t$=500), cp. Fig. \ref{a34} (lower). In the case of ultra-weak coupling ($\kappa = 0.0125$), the oscillators are at the beginning quite well synchronized (for $t<300$), as their behavior is still dominated by the common initial conditions.
Later on, the phase differences are gradually increased, leading to almost uniform distribution for $D(t) > 0.9$.

\begin{figure}[h]
 \includegraphics[angle=270,width=8cm]{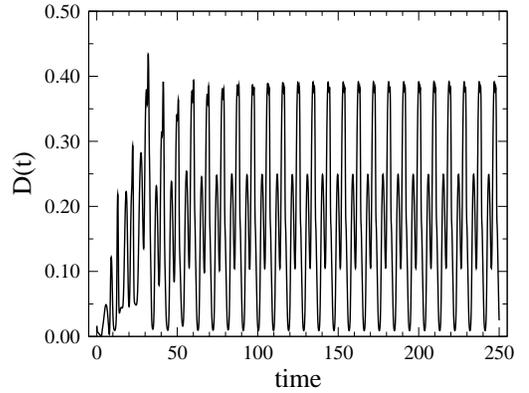}
\caption{Time dependent circular variance $D(t)$ for $N=324$ coupled oscillators
arranged on a cluster of the square lattice with six impurities and $\kappa = 0.5$.}
\label{a7}
\end{figure}

In the case of regular rings, better synchronization is achieved for declustered rings, having shorter average path length, cp. Fig. \ref{a56}. The average variance is smaller, 0.4 for declustered comparing to 0.6 for clustered systems, while the transient period is shorter, $t \approx 100$ for declustered relative to $t \approx 200$ for clustered rings. As reasonable, from the numbers above we see
that the transient time necessary to attain a stationary evolution increases when the coupling constant is decreased.

Finally, as in the main text, we address the question of whether  the average path length is the only topological parameter that matters. To this end, we compare the circular variance for the declustered regular ring of 100 elements with that for a cluster of the square lattice with 324 elements.  Both networks have squares as basic cycles and almost the same average path lengths, while the standard deviation of the distribution of path lengths
is smaller for the square lattice (see Subsec. \ref{ss_cs}). Comparing Figs. \ref{a7} and \ref{a56} (lower) we note that the square network has a smaller average variance ($D \approx 0.2$) and a shorter  transient  ($t \approx 60$). Thus, we conclude again that smaller deviation of path lengths improves synchronization and
shortens the time needed to reach the steady state.


\begin{thebibliography}{99}
\bibitem{note}
Eqs. (1a) and (1b) of Ref. \cite{EL02} contain an errata: the coupling term should
be in Eq. (1a), as here,  and not in Eq. (1b). However, all numerical results presented
therein correspond to the correct writing of the equations shown in this work.
\end{thebibliography}
\end{document}